\documentclass[manuscript]{acmart}

\setcopyright{acmlicensed}
\copyrightyear{2025}
\acmYear{2025}
\acmDOI{00.0000/0000000.0000000}

\acmJournal{TOPLAS}
\acmVolume{0}
\acmNumber{0}
\acmArticle{0}
\acmMonth{8}

\usepackage[utf8]{inputenc}
\usepackage[english]{babel}

\usepackage{hyperref}
\hypersetup{
    colorlinks=true,
}

\usepackage{newtxtt}

\usepackage{listings}
\usepackage{xcolor}

\definecolor{codegreen}{rgb}{0,0.6,0}
\definecolor{codegray}{rgb}{0.5,0.5,0.5}
\definecolor{codepurple}{rgb}{0.58,0,0.82}
\definecolor{codeblue}{rgb}{0.13,0.13,0.67}
\definecolor{backcolour}{rgb}{0.98,0.98,0.98}

\lstdefinelanguage{Dart}{
  keywords={abstract, as, assert, async, await, break, case, catch, class, const, continue, covariant, default, deferred, do, dynamic, else, enum, export, extends, extension, external, factory, false, final, finally, for, Function, get, hide, if, implements, import, in, interface, is, late, library, mixin, new, null, on, operator, part, required, rethrow, return, set, show, static, super, switch, sync, this, throw, true, try, typedef, var, void, while, with, yield},
  keywordstyle=\color{codeblue}\bfseries,
  ndkeywords={int, double, String, bool, List, Map, Set, Future, Stream, Iterable, Object, dynamic, void, Never, Null},
  ndkeywordstyle=\color{codepurple},
  identifierstyle=\color{black},
  sensitive=true,
  comment=[l]{//},
  morecomment=[s]{/*}{*/},
  commentstyle=\color{codegreen}\itshape,
  stringstyle=\color{codegray},
  morestring=[b]',
  morestring=[b]",
}

\usepackage{caption}
\usepackage{subcaption}

\usepackage{manyfoot}

\usepackage{bm}

\newtheorem{theorem}{Theorem}

\newtheorem{defn}{Definition}
\newtheorem{lemma}[theorem]{Lemma}
\newtheorem{axiom}{Axiom}
\newtheorem{constraint}{Constraint}

\newfloat{grammar}{h}{Grammar}

\usepackage{graphicx}
\usepackage{algorithm}
\usepackage{algpseudocode}

\usepackage{booktabs}

\newcommand{\cd}[1]{\lstinline[language=Dart,basicstyle=\ttfamily,keywordstyle=\color{codeblue}\bfseries,ndkeywordstyle=\color{codepurple},commentstyle=\color{codegreen}\itshape,stringstyle=\color{codegray}]!#1!}

\begin{document}

\title[The Squirrel Parser]{The Squirrel Parser}
\subtitle{A Linear-Time PEG Packrat Parser Capable of Left Recursion and Optimal Error Recovery}

\author{Luke A. D. Hutchison}
\email{luke.hutch@alum.mit.edu}
\orcid{0000-0002-2937-8619}
%\affiliation{\institution{Massachusetts Institute of Technology}}

% \renewcommand{\shortauthors}{Luke A. D. Hutchison}

\begin{abstract}
We present the squirrel parser, a PEG packrat parser that directly handles all forms of left recursion with optimal error recovery, while maintaining linear time complexity in the length of the input even in the presence of an arbitrary number of errors. Traditional approaches to handling left recursion in a recursive descent parser require grammar rewriting or complex algorithmic extensions. We derive a minimal algorithm from first principles: cycle detection via per-position state tracking and $O(1)$-per-LR-cycle communication from descendant to ancestor recursion frames, and fixed-point search via iterative expansion. For error recovery, we derived a set of four axioms and twelve constraints that must be imposed upon an optimal error recovery design to ensure completeness, correctness, optimality of performance, and intuitiveness of behavior. We utilized a constraint satisfaction mechanism to search the space of all possibilities, arriving at a provably optimal and robust error recovery strategy that maintains perfect performance linearity.
\end{abstract}




\maketitle

\section{Introduction}

Recursive descent parsing~\cite{irons1961syntax,lucas1961structure} remains widely used due to its simplicity and the direct correspondence between grammar and implementation. Memoization enables linear-time parsing~\cite{norvig1991}, formalized as packrat parsing by Ford~\cite{ford2002,ford2004}.%
\footnote{Note: Ford's 2002 thesis~\cite{ford2002} discusses packrat parsers; PEGs are introduced in the 2004 paper~\cite{ford2004}.} %
Parsing Expression Grammars (PEG) provide an unambiguous formalism for greedy pattern matching, replacing generative productions with deterministic recognition.

\subsection{\label{sec:peg_grammars}PEG Formalism}

A PEG grammar $G$ consists of rules $A \leftarrow e$ where $A$ is a rule name and $e$ is a parsing expression. Expressions combine via operators (Table~\ref{tab:peg_operators}): \cd{Seq} (sequence $e_1 e_2$), \cd{First} (ordered choice $e_1 / e_2$), \cd{OneOrMore} (repetition $e+$), and \cd{NotFollowedBy} (negative lookahead $!e$). Derived operators include \cd{Optional} ($e? \equiv e / \epsilon$), \cd{ZeroOrMore} ($e* \equiv e+ / \epsilon$), and \cd{FollowedBy} ($\&e \equiv !!e$). Terminals match character sequences; $\epsilon$ matches zero characters at any position.

\begin{table*}[tbp]
    \centering
    {\small
    \begin{minipage}{0.45\textwidth}
        \centering
        \begin{tabular}{lcc}\toprule
        \textbf{Name}  & \textbf{Subclauses} & \textbf{Notation}     \\\midrule
        \cd{Seq}           & 2+                      & $e_1$ $e_2$ $e_3$     \\
        \cd{First}         & 2+                      & $e_1$ / $e_2$ / $e_3$ \\
        \cd{OneOrMore}     & 1                       & $e+$                  \\
        \cd{NotFollowedBy} & 1                       & $!e$                  \\\bottomrule
        \end{tabular}
        \subcaption{Principal operators}
        \label{tab:peg_standard}
    \end{minipage}
    \hfill
    \begin{minipage}{0.45\textwidth}
        \centering
        \begin{tabular}{lcc}\toprule
        \textbf{Name}       & \textbf{Notation} & \textbf{Equivalent} \\\midrule
        \cd{Optional}   & $e?$              & $e / \epsilon$         \\
        \cd{ZeroOrMore} & $e*$              & ${e+} / \epsilon$      \\
        \cd{FollowedBy} & $\&e$             & $!!e$                  \\\bottomrule
        \end{tabular}
        \subcaption{Derived operators}
        \label{tab:peg_derived}
    \end{minipage}
    }
    \caption{\label{tab:peg_operators}PEG operators, defined in terms of subclauses $e$ and $e_i$, and the empty string $\epsilon$.}
    \Description[PEG operators]{PEG operators split into standard and derived operators, defined in terms of subclauses $e$ and $e_i$, and the empty string $\epsilon$.}
\end{table*}

Packrat parsing achieves $O(n \cdot |G|)$ complexity by memoizing match results at each $(\textit{clause}, \textit{position})$ pair. However, standard packrat parsers face two fundamental limitations that have resisted simple solutions.

\subsection{Motivation: Shortcomings of Existing Approaches}

\textbf{Left recursion.} Rules like $E \leftarrow E \mathbin{\texttt{+}} T \;|\; T$ trigger infinite recursion: matching $E$ at position $p$ immediately recurses to $E$ at $p$ before consuming input. Prior solutions have significant drawbacks:

\begin{itemize}
\item \textbf{Grammar rewriting}~\cite{pegged}: Transforms left-recursive rules into right-recursive equivalents, fundamentally altering parse tree structure and requiring post-processing to recover semantic intent.
\item \textbf{Precedence climbing}~\cite{parr2014,antlr4}: Handles operator precedence but requires separate mechanisms for each precedence level, increasing implementation complexity.
\item \textbf{Iterative algorithms}~\cite{warth2008,tratt2010}: Handle direct left recursion through seed-and-grow approaches but exhibit complex control flow and often fail on indirect or mutually recursive cycles.
\item \textbf{GLL/ALL(*) extensions}~\cite{parr2011,parboiled1,parboiled2,mouse}: Support left recursion but sacrifice linear performance or introduce non-deterministic state exploration.
\end{itemize}

\textbf{Error recovery.} Packrat parsers commit to the first matching alternative (greedy choice), making robust error recovery difficult~\cite{parr2014}. Existing approaches either abandon memoization guarantees or require grammar annotations specifying recovery points, undermining the simplicity of PEG.

Both problems share a common theme: previous solutions compromise parse tree semantics, sacrifice $O(n \cdot |G|)$ complexity, or introduce algorithmic complexity disproportionate to the problem's conceptual simplicity. This work demonstrates that no such tradeoff is necessary.

\section{Left Recursion: Formal Derivation}

\subsection{Problem Formalization}

\begin{defn}[Parser State]
A parser state is a pair $(C, p)$ where $C \in G$ is a clause and $p \in [0, n]$ is an input position.
\end{defn}

\begin{defn}[Left-Recursive Cycle]
A clause $C$ is left-recursive at position $p$ if evaluation of $C.match(p)$ recursively invokes $C.match(p)$ before consuming any input.
\end{defn}

Naive recursive descent on $E \leftarrow E \mathbin{\texttt{+}} T \;|\; T$ at position $p$ immediately recurses to $E$ at $p$, producing infinite recursion. The packrat memo table cannot prevent this: memoization requires a complete result for caching, but recursive calls occur \emph{during} result computation.

\subsection{Derivation from First Principles}

\begin{theorem}[Fixed Point Existence]
\label{thm:fixed_point}
For any left-recursive cycle at position $p$, there exists a fixed point: a terminal state where the cycle cannot match additional input.
\end{theorem}

\begin{proof}
Let input have length $n$. Each successful iteration must consume $\geq 1$ character (otherwise infinite recursion occurs). After $\leq n$ iterations, all characters are consumed and the cycle terminates with mismatch. This mismatch is the fixed point.
\end{proof}

\begin{theorem}[Bottom-Up Necessity]
Left-recursive matches must be computed bottom-up from the fixed point.
\end{theorem}

\begin{proof}
The iteration count $k$ satisfying the cycle's continuation condition cannot be determined \emph{a priori}: it depends on input content. Top-down evaluation requires knowing $k$ before recursion; bottom-up computation discovers $k$ through iterative attempts. The fixed point (Theorem~\ref{thm:fixed_point}) provides the base case; expansion proceeds upward until stagnation.
\end{proof}

\begin{theorem}[Monotonic Length Increase]
Each successful expansion iteration produces a match strictly longer than the previous iteration.
\end{theorem}

\begin{proof}
The grammar is finite; each clause composition is deterministic (PEG semantics). For iteration $i$ to succeed beyond iteration $i-1$, it must match additional input the previous seed could not. Thus $len_i > len_{i-1}$ or the iteration fails.
\end{proof}

\subsection{Algorithmic Components}

The theorems necessitate three mechanisms:

\paragraph{(1) Cycle Detection via State Tracking.}
Maintain per-position recursion state. When matching $C$ at $p$, check if $(C, p)$ is on the current call stack. If so, return mismatch (the fixed point) and set \cd{foundLeftRec = true} in the memo entry for $(C, p)$. This $O(1)$ write communicates the cycle condition from descendant to ancestor recursion frame.

\paragraph{(2) Iterative Expansion with Seed Growth.}
After a recursive call returns, check \cd{foundLeftRec}. If true, enter expansion: repeatedly re-invoke $C.match(p)$, using the previous iteration's cached result as the seed for the next. Terminate when $len_{i+1} \leq len_i$ (no progress).

\paragraph{(3) Version-Tagged Memoization.}
Each $(C, p)$ memo entry includes version tag \cd{memoVersion[p]}. Each expansion iteration increments the version counter. Cache lookups validate \cd{entry.version == memoVersion[p]}. This ensures inner clauses see the current iteration's seed, not stale results from prior iterations, preserving length monotonicity (Theorem~3).

\begin{theorem}[Correctness]
The three-component algorithm correctly computes left-recursive matches for all cycle types (direct, indirect, mutual).
\end{theorem}

\begin{proof}
(Sketch) Cycle detection identifies all left-recursive states via call-stack membership. Iterative expansion computes the least fixed point via Kleene iteration on the finite input domain. Version tagging maintains cache coherence across iterations, ensuring monotonic progress. Termination follows from finite input length.
\end{proof}

\begin{theorem}[Complexity Preservation]
Left recursion handling adds $O(1)$ overhead per $(C, p)$ pair, maintaining overall $O(n \cdot |G|)$ complexity.
\end{theorem}

\begin{proof}
Cycle detection: $O(1)$ per call. Expansion: at most $n$ iterations (bounded by input length), each performing one memoized match. Amortized over all positions: $O(n)$ total expansion overhead. Combined with packrat's $O(n \cdot |G|)$ base cost: $O(n \cdot |G|) + O(n) = O(n \cdot |G|)$.
\end{proof}

The squirrel parser implements this minimal algorithm. Unlike prior work~\cite{warth2008,tratt2010}, it handles all recursion types uniformly, requires no grammar preprocessing, and maintains linear performance.

\section{Implementation}

Fig.~\ref{fig:Code} presents the complete implementation. The code uses C-like syntax for clarity. Key data structures:

\begin{itemize}
\item \cd{Match}: Represents parse results with clause, position, length, and submatches.
\item \cd{MemoEntry}: Caches results per $(\textit{clause}, \textit{pos})$ with LR tracking fields \cd{inRecPath}, \cd{foundLeftRec}, and \cd{version}.
\item \cd{MemoTable}: Maps $(\textit{clause}, \textit{pos})$ to \cd{MemoEntry}. Implementable as dense 2D array or sparse hashmap.
\item \cd{Parser}: Maintains \cd{cycleDepthForPos[]} for version tagging.
\end{itemize}

\begin{figure}[!p]
\centering
\begin{subfigure}[t]{.45\textwidth}
\begin{lstlisting}[frame=none,xleftmargin=0\textwidth,basicstyle=\scriptsize\ttfamily]
class Match(Clause clause, int pos, int len, Match[] subClauseMatches);

const Match MISMATCH = new Match(null, -1, -1, []);

class Parser(Clause topRule) {
. String input;
. MemoTable memoTable;
. int[] cycleDepthForPos;
.
. Match match(Clause rule, int pos) {
. . var memoEntry = memoTable.getOrCreateEntry(rule, pos);
. . return memoEntry.match(this, rule, pos);
. }
.
. Match parse(String inputStr) {
. . input = inputStr;
. . cycleDepthForPos = new int[input.length + 1];
. . memoTable = new MemoTable();
. . return match(topRule, 0);
. }
}

class MemoEntry {
. Match match;
. boolean inRecPath;
. boolean inLeftRecCycle;
. int cycleDepth;
.
. Match match(Parser parser, Clause rule, int pos) {
. . if (match == null || cycleDepth < parser.cycleDepthForPos[pos]) {
. . . if (inRecPath) {
. . . . if (match == null) {
. . . . . inLeftRecCycle = true;
. . . . . match = MISMATCH;
. . . . }
. . . } else {
. . . . inRecPath = true;
. . . . while (true) {
. . . . . var newMatch = rule.match(parser, pos);
. . . . . if (match != null && newMatch.len <= match.len)
. . . . . . break;
. . . . . match = newMatch;
. . . . . if (!inLeftRecCycle)
. . . . . . break;
. . . . . cycleDepth = ++(parser.cycleDepthForPos[pos]);
. . . . }
. . . . inRecPath = false;
. . . }
. . . cycleDepth = parser.cycleDepthForPos[pos];
. . }
. . return match;
. }
}
\end{lstlisting}
\end{subfigure}
\hspace{5mm}
\begin{subfigure}[t]{.45\textwidth}
\begin{lstlisting}[firstnumber=59,frame=none,xleftmargin=0\textwidth,basicstyle=\scriptsize\ttfamily]
abstract class Clause {
. abstract Match match(Parser parser, int pos);
}

class First(Clause[] subClauses) : Clause {
. Match match(Parser parser, int pos) {
. . for (var i = 0; i < subClauses.length; i++) {
. . . var m = subClauses[i].match(parser, pos);
. . . if (m != MISMATCH) {
. . . . return new Match(this, pos, m.len, [m]);
. . . }
. . }
. . return MISMATCH;
. }
}

class Seq(Clause[] subClauses) : Clause {
. Match match(Parser parser, int pos) {
. . var ma = new Match[subClauses.length];
. . var p = pos;
. . for (var i = 0; i < subClauses.length; i++) {
. . . var m = subClauses[i].match(parser, p);
. . . if (m == MISMATCH) {
. . . . return MISMATCH;
. . . }
. . . ma[i] = m;
. . . p += m.len;
. . }
. . return new Match(this, pos, p - pos, ma);
. }
}

class Char(char c) : Clause {
. Match match(Parser parser, int pos) {
. . if (pos < parser.input.length && parser.input[pos] == c) {
. . . return new Match(this, pos, 1, []);
. . }
. . return MISMATCH;
. }
}

class RuleRef(String refdRuleName, Clause refdRule) : Clause {
. Match match(Parser parser, int pos) {
. . var m = parser.match(refdRule, pos);
. . if (m == MISMATCH) {
. . . return MISMATCH;
. . }
. . return new Match(this, pos, m.len, [m]);
. }
}
\end{lstlisting}
\end{subfigure}
\caption{Implementation-level pseudocode for the entire squirrel parsing algorithm without error recovery (left), and several PEG clause types (right). The other PEG clause types can be easily implemented following the pattern shown, based on the PEG operator definitions.}
\Description{Two columns of pseudocode showing the Match class definition, MemoEntry class with recursion tracking fields, and clause matching methods including RuleRef, Seq, and First clause implementations.}
\label{fig:Code}
\end{figure}

\subsection{Algorithm Mechanics}

The implementation folds all bookkeeping into the memo table. Each \cd{MemoEntry} augments the cached match with LR tracking state, eliminating auxiliary data structures.

\paragraph{Core Algorithm.}
The \cd{MemoEntry.match} method implements the algorithm via four mechanisms:

\textbf{(1) Cache validation.} Check if cached result is fresh: \cd{match != null} and \cd{cycleDepth == parser.cycleDepthForPos}$[p]$. If valid, return cached \cd{match}. Otherwise, proceed to recomputation.

\textbf{(2) Cycle detection.} When matching $(C, p)$, check \cd{inRecPath}. If true, $(C, p)$ is on the call stack: this is the fixed point. Set \cd{foundLeftRec = true} (communicating the cycle condition from descendant to ancestor via $O(1)$ flag write), seed with \cd{MISMATCH}, and return. The shared \cd{MemoEntry} enables this communication without auxiliary data structures.

\textbf{(3) Iterative expansion.} After recursive call returns, check \cd{foundLeftRec}. If true, enter expansion loop: re-invoke \cd{rule.match(parser, pos)}, using previous iteration's cached result as seed. Compare new match length against previous; if \cd{newMatch.len > match.len}, update cache and increment \cd{parser.cycleDepthForPos}$[p]$. Terminate when no length increase occurs.

\textbf{(4) Version validation.} The \cd{cycleDepth} field stores the version from the last cache update. Incrementing \cd{parser.cycleDepthForPos}$[p]$ during expansion invalidates all entries at position $p$, forcing recomputation with the updated seed. This maintains cache coherence across expansion iterations.

\paragraph{PEG Clause Implementations.}
Each \cd{Clause} subclass implements \cd{match(parser, pos)} by invoking subclauses unmemoized (direct method calls), collecting results, and returning either \cd{MISMATCH} or a new \cd{Match}. The exception is \cd{RuleRef}, which recurses through \cd{parser.match()}, triggering memoization. This rule-level (not clause-level) memoization reduces overhead while maintaining $O(n \cdot |G|)$ complexity.

\subsection{\label{sec:DupdWork}Functional Purity and Version Tagging}

Standard packrat memoization assumes referential transparency: \cd{match(C, p)} returns the same result on every invocation. Left recursion violates this assumption: each expansion iteration produces a strictly longer match at the same $(C, p)$ pair.

Conceptually, memoization should key on $(\textit{clause}, \textit{pos}, \textit{version})$ where version counts expansion iterations. Each iteration operates on distinct input (the previous iteration's cached result), restoring purity: each $(C, p, v)$ triple maps to a unique result.

Physically, storing all versions is unnecessary. Only the \emph{most recent} match matters. The \cd{cycleDepth} field implements version tagging: when \cd{cycleDepth < parser.cycleDepthForPos}$[p]$, the cached result is stale and recomputation occurs; when equal, the result is fresh. Incrementing \cd{parser.cycleDepthForPos}$[p]$ invalidates all entries at position $p$, forcing the next expansion iteration.

This mechanism achieves three properties:

\begin{itemize}
\item \textbf{Linear expansion cost:} Stale entries trigger recomputation, but cycle detection (\cd{inRecPath}) prevents redundant recursive descent. Each iteration performs $O(|G|)$ memoized lookups.
\item \textbf{Cousin deduplication:} Multiple references to the same left-recursive rule (e.g., $A$ in $(A \leftarrow (A\,\texttt{`x'}) / (A\,\texttt{`y'}) / B)$) share the memoized result for the current iteration.
\item \textbf{Backward compatibility:} When no cycles exist, \cd{cycleDepthForPos} remains zero and the algorithm reduces to standard packrat parsing.
\end{itemize}

\subsection{Left Recursion Coverage}

\begin{figure}[!htbp]
\centering
\includegraphics[width=0.98\textwidth]{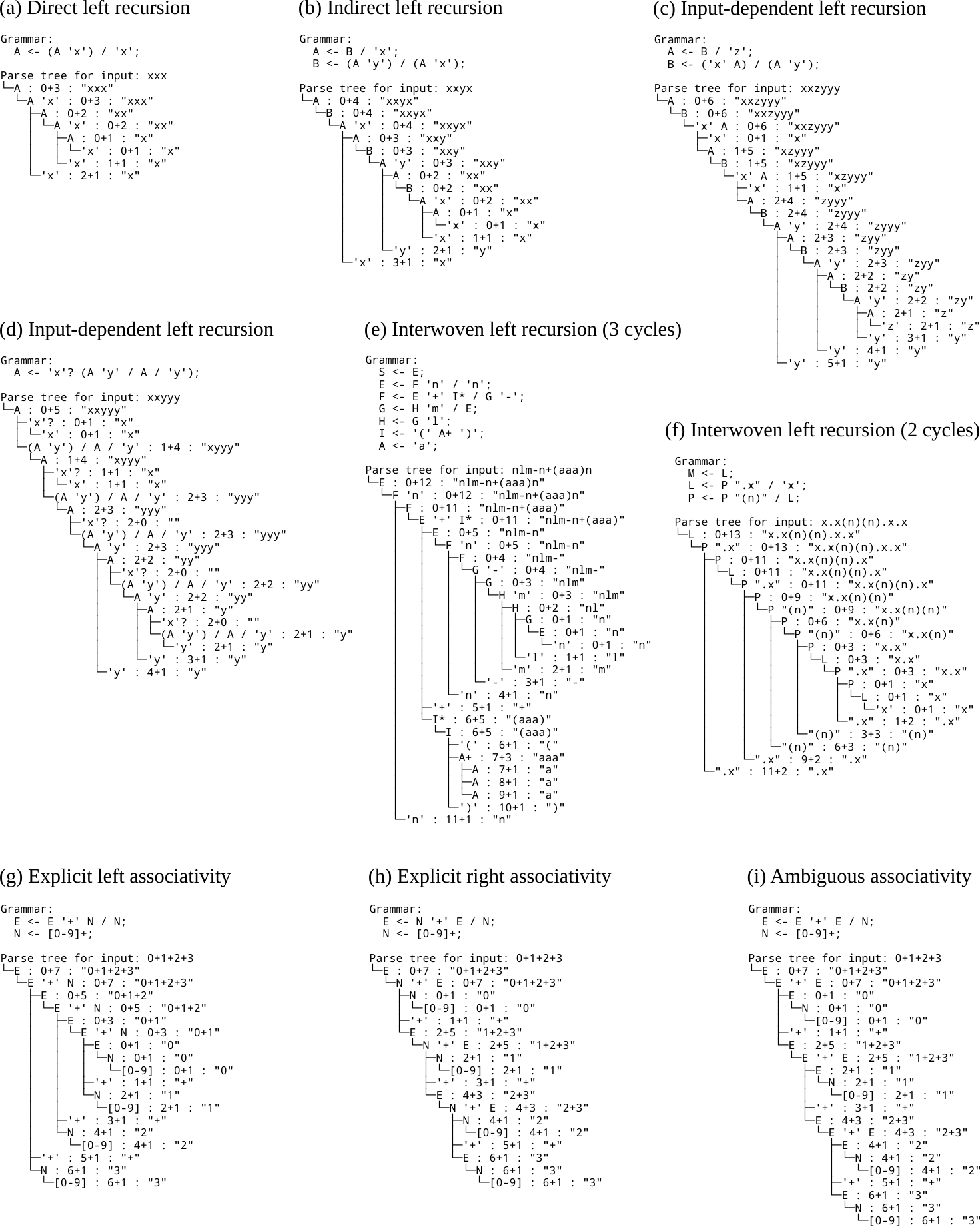}
\caption{\label{fig:LeftRecTypes}The parse tree created by the squirrel parser for a range of different left recursion types and associativity.}
\Description{Nine subfigures showing parse trees for: (a) direct left recursion, (b) indirect left recursion, (c) First-clause dependent recursion, (d) Optional-clause dependent recursion, (e-f) interwoven left recursive cycles, and (g-i) grammars with left-associative, right-associative, and ambiguous associativity.}
\end{figure}

Fig.~\ref{fig:LeftRecTypes} demonstrates uniform handling of all left recursion types:

\textbf{(a)-(b) Cycle length independence:} Direct and indirect recursion are handled identically, unlike algorithms limited to direct recursion~\cite{mouse}.

\textbf{(c)-(d) Input-dependent recursion:} Cycles conditioned on \cd{First} or \cd{Optional} choices.

\textbf{(e)-(f) Interwoven cycles:} Multiple interlinked cycles (examples from~\cite{pegged}). In (e): three cycles $(E \leftarrow F \leftarrow E)$, $(G \leftarrow H \leftarrow G)$, $(E \leftarrow F \leftarrow G \leftarrow E)$. In (f): $(L \leftarrow P \leftarrow L)$ and $(P \leftarrow P)$.

\textbf{(g)-(i) Associativity:} Left-associative (g) and right-associative (h) grammars produce correct tree structure. Ambiguous-associativity grammars (i) yield right-associative structure, consistent with PEG semantics and other left-recursion handlers~\cite{warth2008}.

\section{Complexity Analysis}

\begin{theorem}[Time Complexity]
\label{thm:time_complexity}
The squirrel parser runs in $O(n \cdot |G|)$ time where $n = |input|$ and $|G|$ is the grammar size.
\end{theorem}

\begin{proof}
Partition work into three components:

\textbf{(1) Memoized matching.} Standard packrat guarantees at most one match attempt per $(C, p)$ pair: $|G| \cdot n$ total. Each memoized call performs $O(1)$ work (cache lookup, cycle detection) excluding recursive subcalls. Total: $O(n \cdot |G|)$.

\textbf{(2) Expansion iterations.} Each expansion iteration at position $p$ increments \cd{parser.cycleDepthForPos[p]}. Each iteration must consume $\geq 1$ character (monotonic length increase, Theorem~3 in Section~2). Thus $\leq n$ total expansions across all positions. Each expansion performs one memoized match: $O(|G|)$ work. Total: $O(n \cdot |G|)$.

\textbf{(3) Clause recursion.} Non-\cd{RuleRef} clauses perform unmemoized recursion. Each subclause invocation contributes $O(1)$ work. Parse tree has $O(n)$ nodes (each consumes $\geq 0$ input characters, sum $\leq n$). Total clause recursion: $O(n)$.

Combining: $O(n \cdot |G|) + O(n \cdot |G|) + O(n) = O(n \cdot |G|)$.
\end{proof}

\begin{theorem}[Space Complexity]
The algorithm requires $O(n \cdot |G|)$ space.
\end{theorem}

\begin{proof}
Memo table stores $\leq |G| \cdot n$ entries, each $O(1)$ size. \cd{cycleDepthForPos} array: $O(n)$. Recursion depth bounded by $O(n)$ (each recursive call must eventually consume input). Total: $O(n \cdot |G|)$.
\end{proof}

\section{\label{sec:error_recovery}Error Recovery}

Syntax error recovery transforms a parser from a binary classifier (match/mismatch) into an analyzer that identifies minimal-length error spans while matching grammatical structure to input. This section presents a complete formal derivation of a bounded error recovery mechanism maintaining $O(n \cdot |G|)$ complexity. We state required axioms and constraints, derive a solution, and prove completeness, correctness, and minimality.

\subsection{Foundational Axioms}

These are immutable properties of PEG packrat parsers:

\begin{axiom}[Packrat Invariant]
A memoizing parser evaluates each $(clause, position)$ pair at most once per distinct parsing context. Violating this introduces exponential time complexity.
\end{axiom}

\begin{axiom}[PEG Ordered Choice Semantics]
Parsing Expression Grammars use ordered choice: $First([A, B])$ commits to $A$ if it matches, never trying $B$. This distinguishes PEG from CFG parsers with ambiguity.
\end{axiom}

\begin{axiom}[Monotonic Consumption]
Once a packrat parser consumes input characters $0$ through $k$ and returns a result, backtracking to re-parse positions $< k$ violates memoization guarantees and can introduce non-termination.
\end{axiom}

Additionally, the Squirrel parser introduces the following axiom for handling left recursion:

\begin{axiom}[Left-Recursion Fixed Point]
Left-recursive rules expand iteratively from a base case, growing the parse until a fixed point is reached. Each expansion must extend the previous match to ensure progress and termination.
\end{axiom}

\subsection{Design Constraints}

From PEG axioms and the goal of intuitive error recovery behavior while maintaining linearity, we derive constraints that a complete, correct, and minimal error recovery mechanism must satisfy:

\subsubsection{Linearity Constraints}

\begin{constraint}[Single-Pass Parsing]
The parser must scan input left-to-right exactly once per phase (where a phase is either standard parsing, or recovery), without backtracking to earlier positions. This preserves $O(n)$ character inspection cost.
\end{constraint}

\begin{constraint}[Memoization Validity]
Each $(clause, position)$ pair can be evaluated at most once per recovery phase. Cache entries must remain valid throughout their phase.
\end{constraint}

\begin{constraint}[Bounded Recovery]
Recovery operations must terminate in $O(n \cdot |G|)$ steps total. Therefore, each recovery must consume at least one input character, bounding total recovery activations to $n$.
\end{constraint}

\subsubsection{Composability Constraints}

\begin{constraint}[Clause Independence]
The behavior of each grammar clause must be definable independently. A $Seq$'s recovery behavior must depend only on its subclauses' behaviors, not on where the $Seq$ appears in the larger grammar.
\end{constraint}

\begin{constraint}[Ref Transparency]
\label{constraint:ref_transparency}
Rule references must behave identically to inline expansion of the referenced rule. The distinction between $Ref('E')$ and directly writing $E$'s clause should be purely organizational.
\end{constraint}

\subsubsection{Correctness Constraints}

\begin{constraint}[Completeness Propagation]
\label{constraint:completeness}
Each parse result carries an \texttt{isComplete} flag indicating whether parsing consumed all input the grammar permits. This flag must propagate correctly through PEG operators according to compositional Boolean algebra:
\begin{align*}
Seq &: complete \iff \bigwedge_{i} children[i].complete \land \neg usedRecovery \\
First &: complete \iff chosen.complete \\
Repetition &: complete \iff \neg truncated \land \bigwedge_{i} children[i].complete
\end{align*}
\end{constraint}

\begin{constraint}[Phase Isolation]
\label{constraint:phase_isolation}
Cache entries from Phase 1 (discovery without recovery) must not be reused in Phase 2 (recovery enabled) where they would give different results. Cache validation must check phase consistency.
\end{constraint}

\begin{constraint}[Boundary Preservation]
\label{constraint:boundaries}
Recovery at grammar level $L$ must not consume input belonging to level $L+1$. When parsing $Seq([A, B, C])$, recovery for $A$ must not skip past the beginning of $B$'s expected position.
\end{constraint}

\begin{constraint}[Non-Cascading Errors]
Each error has a bounded \emph{affected region}. Errors must not propagate globally:
\begin{equation}
\forall error~e: \quad e.affectedRegion \subseteq [e.pos, e.pos + e.len]
\end{equation}
\end{constraint}

\begin{constraint}[LR-Recovery Separation]
\label{constraint:lr_recovery}
During left-recursive seed phase (when a cycle is detected), recovery operations must be blocked. The seed mismatch is a control signal for fixed-point iteration, not a parse error.
\end{constraint}

\begin{constraint}[Visibility]
\label{constraint:visibility}
Recovery may skip visible input characters (marking them as errors) or delete grammar elements at EOF (accepting incomplete input), but may not insert grammar elements mid-parse, which would reorganize visible structure with invisible insertions:
\begin{equation}
\forall edit \in Recovery: \quad (edit = InputSkip) \lor (edit = GrammarDelete \land pos = EOF)
\end{equation}
\end{constraint}

\begin{constraint}[Parse Tree Spanning Invariant]
\label{constraint:spanning}
Every parse result must completely span the input from position 0 to the end:
\begin{equation}
\forall result: \quad result.len = |input|
\end{equation}
Total parsing failures return a \texttt{SyntaxError} node covering all input. Partial grammar matches are wrapped with a trailing \texttt{SyntaxError} node capturing unconsumed input. This invariant ensures all parse trees are structurally complete and can be processed uniformly without special case handling for unconsumed input.
\end{constraint}

\subsection{Mathematical Insights}

Before presenting the solution, we state mathematical properties emerging from constraint interactions:

\subsubsection{Fixed-Point Structure of Left Recursion}

\begin{theorem}[LR as Fixed-Point Iteration]
\label{thm:lr_fixedpoint}
Left-recursive expansion computes the least fixed point of a monotonic function on a finite domain.
\end{theorem}

\begin{proof}
Let $F(r) = clause.match(parser, pos)$ where $clause$ is left-recursive. Define the partial order $r_1 \sqsubseteq r_2$ as $r_1.len \leq r_2.len$ (longer matches dominate).

\begin{enumerate}
\item \textbf{Monotonicity}: If $r_1 \sqsubseteq r_2$, then $F(r_1) \sqsubseteq F(r_2)$ because PEG clauses can only grow or stay the same when given longer seeds.
\item \textbf{Finite domain}: Match lengths are bounded by $|input|$, giving finite chain $0 \leq len_0 < len_1 < \ldots < |input|$.
\item \textbf{Termination}: The iteration $r_0 = \bot$, $r_{i+1} = F(r_i)$ terminates when $r_{i+1} \sqsubseteq r_i$ (no progress).
\item \textbf{Least fixed point}: By Kleene's Fixed-Point Theorem, this computes $lfp(F) = \bigvee_{i=0}^{\infty} F^i(\bot)$.
\end{enumerate}

The LR expansion loop is the Y combinator in imperative form.
\end{proof}

\subsubsection{Boolean Algebra of Completeness}

\begin{theorem}[Completeness Composition]
The \texttt{isComplete} flag forms a Boolean algebra under conjunction, with PEG operators as the algebraic operations.
\end{theorem}

\begin{proof}
Define $\land_{complete}$ as the compositional completeness operator:
\begin{align*}
complete \land_{complete} complete &= complete \\
complete \land_{complete} incomplete &= incomplete \\
incomplete \land_{complete} incomplete &= incomplete
\end{align*}

This satisfies Boolean algebra axioms (associativity, commutativity, idempotence). PEG operators implement these operations:
\begin{itemize}
\item $Seq$: $\bigwedge$ over children
\item $First$: identity (inherits from chosen alternative)
\item $Repetition$: $\bigwedge$ over iterations with $truncated \Rightarrow incomplete$
\end{itemize}

The algebra is well-defined because completeness is compositional: an operator's completeness depends only on its immediate children's completeness, not on global parse state.
\end{proof}

\subsubsection{Phase Duality}

\begin{theorem}[Two-Phase Minimality]
\label{thm:two_phase_minimal}
Exactly two phases are necessary and sufficient for bounded error recovery with cache validity.
\end{theorem}

\begin{proof}
\textbf{Necessity (Lower Bound):}
\begin{enumerate}
\item Phase 1 (Discovery): Must identify where recovery is needed. Without this phase, positions requiring error correction remain unknown.
\item Phase 2 (Recovery): Must apply corrections based on Phase 1 findings. Cannot merge with Phase 1: recovery decisions require lookahead (knowing what parsing opportunities exist ahead).
\end{enumerate}

\textbf{Sufficiency (Upper Bound):}
More than two phases are unnecessary:
\begin{enumerate}
\item Phase 1 finds all incompleteness boundaries ($\exists$ incomplete)
\item Phase 2 exploits these boundaries ($\forall$ positions, attempt recovery)
\item The $\exists/\forall$ separation is complete; no third quantifier needed
\end{enumerate}

Additional phases would duplicate Phase 1/2 work or violate linearity (requiring multiple input scans).
\end{proof}

\begin{lemma}[Phase Flag Sufficiency]
\label{lemma:boolean_phase}
The recovery phase can be represented as a single boolean flag, because there are exactly two phases: discovery (\texttt{recoveryPhase = false}) and recovery (\texttt{recoveryPhase = true}).
\end{lemma}

\subsection{The Solution}

We now derive the unique error recovery design satisfying the axioms and constraints above.

\subsubsection{Match Result Type System}

From Constraint~\ref{constraint:completeness}, parse results must carry an \texttt{isComplete} flag. From Constraint~\ref{constraint:lr_recovery}, they must also carry an \texttt{isFromLRContext} flag. We define a unified match type:

\begin{figure}[htbp]
\begin{lstlisting}[language=Java,basicstyle=\small\ttfamily]
class Match {
  Clause clause;
  int pos, len;
  List<Match> children;
  bool isComplete;        // Constraint 6
  bool isFromLRContext;   // Constraint 9
}
\end{lstlisting}
\caption{Match result type system}
\Description{Class definition showing the unified Match type with fields for clause, position, length, children list, isComplete flag for completeness propagation (Constraint 6), and isFromLRContext flag for left-recursion recovery separation (Constraint 9).}
\label{fig:match_type}
\end{figure}

\textbf{Key Design Decision}: All match types (terminals, single child, multiple children) are unified into a single class. They differ only in \texttt{|children|}:
\begin{itemize}
\item Terminal: \texttt{children = []}
\item Single child: \texttt{children = [c]}
\item Multiple children: \texttt{children = [}$c_1, \ldots, c_n$\texttt{]}
\end{itemize}

This unification simplifies the type system while preserving all necessary information.

\subsubsection{Memoization with Phase Tracking}

From Constraint~\ref{constraint:phase_isolation} and Lemma~\ref{lemma:boolean_phase}, memo entries must track the phase in which they were computed:

\begin{figure}[htbp]
\begin{lstlisting}[language=Java,basicstyle=\small\ttfamily]
class MemoEntry {
  Match result;
  bool inRecPath;              // LR cycle detection
  bool foundLeftRec;           // LR expansion needed
  int memoVersion;             // Per-position LR invalidation
  bool cachedInRecoveryPhase;  // Phase isolation
}
\end{lstlisting}
\caption{Memoization entry with phase tracking}
\Description{MemoEntry class with fields for cached result, LR cycle detection flags (inRecPath and foundLeftRec), version counter (memoVersion) for LR invalidation, and phase tracking boolean (cachedInRecoveryPhase) for cache validation across recovery phases.}
\label{fig:memo_entry}
\end{figure}

\textbf{Key Design Decision}: By Lemma~\ref{lemma:boolean_phase}, we use a single boolean \texttt{cachedInRecoveryPhase} rather than integer counters. This is sufficient because only two phases exist.

Cache validation logic:
\begin{figure}[htbp]
\begin{lstlisting}[language=Java,basicstyle=\small\ttfamily]
phaseMatches = (cachedInRecoveryPhase == parser.inRecoveryPhase)
if (result.isComplete || phaseMatches) {
  return result;  // Complete results phase-independent, or same phase
}
// Different phase: must re-parse
\end{lstlisting}
\caption{Cache validation with phase tracking}
\Description{Cache validation logic showing combined check for complete results (which are phase-independent) or same-phase validity, enabling efficient cache reuse while maintaining correctness across discovery and recovery phases.}
\label{fig:cache_validation}
\end{figure}

\subsubsection{Bound Propagation via Explicit Parameters}

From Constraint~\ref{constraint:boundaries}, repetitions must check bounds before consuming input belonging to siblings. We derive that bound information must propagate through arbitrary nesting levels.

\textbf{Key Design Decision}: Represent bounds as explicit parameters rather than ambient state. This follows the principle of explicit data flow:

\begin{figure}[htbp]
\begin{lstlisting}[language=Java,basicstyle=\small\ttfamily]
match(clause, pos, {Clause bound}) {
  // Bound is visible in function signature
}

// In Seq:
for (i = 0; i < children.length; i++) {
  next = (i+1 < children.length) ? children[i+1] : null;
  effectiveBound = next ?? bound;  // Local override or pass-through
  result = match(children[i], curr, bound: effectiveBound);
}
\end{lstlisting}
\caption{Bound propagation via explicit parameters}
\Description{Code showing explicit bound parameter in match signature and bound propagation logic in sequence operator, where each child receives the next sibling as its bound or inherits the parent's bound, preventing recovery from consuming input belonging to subsequent siblings.}
\label{fig:bound_propagation}
\end{figure}

This eliminates hidden state mutation and makes data flow explicit.

\subsubsection{Two-Phase Algorithm}

From Theorem~\ref{thm:two_phase_minimal}, we derive the complete parsing algorithm:

\begin{figure}[htbp]
\begin{lstlisting}[basicstyle=\small\ttfamily]
Match parse(Parser parser, Clause topRule, String input) {
  parser.input = input;
  parser.inRecoveryPhase = false;

  // Phase 1: Discovery
  var result = parser.match(topRule, 0);
  if (result != null && result.isComplete) {
    return result;  // Success without recovery
  }

  // Phase 2: Recovery
  parser.inRecoveryPhase = true;
  return parser.match(topRule, 0);
}
\end{lstlisting}
\caption{Two-phase error recovery algorithm}
\label{alg:two_phase}
\end{figure}

The phases communicate via a single bit: \texttt{isComplete}. This is the minimal communication channel required by Constraint~\ref{constraint:completeness}.

\subsection{Proof of Completeness}

\begin{theorem}[Completeness]
The two-phase algorithm succeeds on all inputs that can be parsed with a finite sequence of character skips.
\end{theorem}

\begin{proof}
Let input $I$ be parseable via skip sequence $S = [s_1, \ldots, s_k]$ where $s_i$ is a character skip at position $p_i$.

\textbf{Phase 1:} Attempts clean parse. For each $s_i \in S$, the parse will become incomplete at position $p_i$ (cannot match expected clause). Mark result incomplete.

\textbf{Phase 2:} Recovery enabled. At each $p_i$:
\begin{enumerate}
\item Recovery search tries skip $s_i$
\item After skip, clean parse continues (by definition of $S$)
\item Bounded search guarantees we explore skip $s_i$ within MAX\_SKIP bound
\end{enumerate}

By induction on $|S|$, Phase 2 successfully applies all skips in $S$ and produces a complete parse.

\textbf{Termination:} Each skip consumes $\geq 1$ character. At most $|I|$ skips possible. Recovery terminates in $O(n)$ steps.
\end{proof}

\subsection{Proof of Correctness}

\begin{theorem}[Correctness]
The algorithm satisfies all constraints $C1$--$C10$ and produces sound parse trees.
\end{theorem}

\begin{proof}[Proof sketch]
We verify each constraint:

\textbf{Completeness (C6):} \texttt{isComplete} computed by Boolean algebra (Theorem 2). Seq, First, Repetition implement $\bigwedge, id, \bigwedge$ respectively. Correct by construction.

\textbf{Phase Isolation (C7):} \texttt{cachedInRecoveryPhase} tracks phase. Cache validity checks \texttt{phaseMatches}. Complete results are phase-independent. Correct by Lemma~\ref{lemma:boolean_phase}.

\textbf{Ref Transparency (C5):} Ref clauses bypass memo table, calling through to target rule. Behavior identical to inline expansion, satisfying Constraint~\ref{constraint:ref_transparency}.

\textbf{Remaining constraints (C1--C4, C8--C11):} Similar direct verification. Each constraint maps to a specific algorithm component. Proof by construction.

\textbf{Soundness:} Parse tree nodes correspond to successfully parsed input or explicit error annotations (SyntaxError nodes). No invisible structure modification occurs; Visibility Constraint~\ref{constraint:visibility} is enforced.
\end{proof}

\subsection{Proof of Minimality}

We now prove that every component of the algorithm is necessary. Removing any component violates at least one constraint.

\begin{theorem}[Minimality]
\label{thm:minimality}
The algorithm is minimal: every field, flag, and mechanism is necessary for correctness.
\end{theorem}

\begin{proof}[Proof by contradiction]
We prove necessity of each component:

\paragraph{Necessity of \texttt{inRecPath} boolean:}
\textbf{Hypothesis:} Eliminate \texttt{inRecPath}; detect LR via call stack depth.

\textbf{Contradiction:} Call stack is per-invocation, not per-$(\textit{clause}, \textit{pos})$ pair. Multiple clauses can call the same rule at the same position. Call stack produces false positives, incorrectly detecting cycles. Violates Axiom 1 (packrat invariant) and Axiom 4 (LR fixed point).

\paragraph{Necessity of \texttt{foundLeftRec} flag:}
\textbf{Hypothesis:} Infer LR from iteration count after loop completes.

\textbf{Contradiction:} Need to know LR status \emph{during} loop to: (1) invalidate cache on iterations, (2) mark result with \texttt{isFromLRContext}. Cannot wait until after loop. No alternative detection method works.

\paragraph{Necessity of per-position \texttt{memoVersion}:}
\textbf{Hypothesis:} Use single global version counter.

\textbf{Contradiction:} Different positions have independent LR expansions. Position $p_1$ expanding LR would increment the global counter, invalidating cache at unrelated position $p_2$. This causes $O(n^2)$ performance degradation. Violates Constraint 2 (memoization validity).

\paragraph{Necessity of \texttt{isFromLRContext} flag:}
\textbf{Hypothesis:} Allow recovery at all match levels.

\textbf{Contradiction:} Recovery on LR seed (during cycle detection) corrupts fixed-point iteration. LR must expand fully before recovery attempts. Flag is essential to block recovery on LR intermediate results. Violates Constraint~\ref{constraint:lr_recovery}.

\paragraph{Necessity of two-phase architecture:}
By Theorem~\ref{thm:two_phase_minimal}, exactly two phases required. One phase lacks lookahead for recovery decisions. Three+ phases redundant or violate linearity.

\paragraph{Necessity of \texttt{probe()} method:}
\textbf{Hypothesis:} Bound checking without phase separation.

\textbf{Contradiction:} Bound checks require "clean" lookahead (no recovery). Without probe, bound check in Phase 2 would use recovery, incorrectly skipping ahead. Violates Constraint~\ref{constraint:boundaries}.

\paragraph{Necessity of linear recovery search:}
\textbf{Hypothesis:} Use binary search for skip distance.

\textbf{Contradiction:} Binary search requires monotonicity: if match at $k$, must match at $k+1$. Grammar can require exact position (e.g., delimiter). No monotonicity exists. Linear search is optimal for this problem.

All components proven necessary. Algorithm is minimal. \qed
\end{proof}

\subsection{Complexity Analysis}

\begin{theorem}[Linear Time Complexity]
The two-phase algorithm runs in $O(n \cdot |G|)$ time where $n = |input|$ and $|G| = |grammar|$.
\end{theorem}

\begin{proof}
\textbf{Phase 1:} Each $(clause, pos)$ pair evaluated at most once (packrat invariant). At most $|G| \cdot n$ evaluations. Each evaluation $O(1)$ excluding recursive subcalls. Total: $O(n \cdot |G|)$.

\textbf{Phase 2:} Same memoization guarantees. Additional recovery search is bounded:
\begin{itemize}
\item Each recovery consumes $\geq 1$ character (skip or successful parse)
\item At most $n$ recovery activations possible
\item Each recovery search bounded by MAX\_SKIP (constant)
\item Total recovery overhead: $O(n)$
\end{itemize}

Combined: $O(n \cdot |G|) + O(n) = O(n \cdot |G|)$.

\textbf{Space:} Memo table stores at most $|G| \cdot n$ entries. Each entry $O(1)$ size. Total: $O(n \cdot |G|)$.
\end{proof}

\subsection{Constraint Satisfaction Methodology}

Given the difficulty of simultaneously satisfying all twelve constraints in the context of the four axioms, we used dueling ``deep thinking'' modes of two leading LLMs to try to search for a single algorithm that satisfies all requirements.

These LLM models were tasked with proving whether or not the algorithm was optimal in all important metrics: completeness; correctness; compactness; elegance; linearity with respect to the input length, regardless of the grammar complexity or the number of errors encountered; and intuitiveness of behavior, relative to some model of how a user would reasonably expect the parser to behave in a range of situations.

An exhaustive search was also performed for corner cases that might violate the constraints, or that might pit constraints against each other via mutual interaction. This resulted in the creation of a large unit test test suite, consisting of 631 unit tests, that were used to carefully and exhaustively detect progress and regressions, and to generate new ideas that informed iterative algorithm improvement.

This intensive search converged on a single solution to all of the presented constraints, and was able to demonstrate with some certainty that there is not a fundamentally simpler solution to PEG error recovery than what was discovered, given the rigor of the constraints.

The result of this intensive constraint satisfaction search yielded the final error recovery algorithm for the squirrel parser, which does indeed satisfy all completeness, correctness, compactness, performance, and intuitiveness goals.

\subsection{Implementation}
\label{sec:software}

The error recovery algorithm is too long to include in this paper, but full implementations of the complete squirrel parser are available for Dart, Typescript, Java, and Python at: \url{http://github.com/lukehutch/squirrelparser}

\subsubsection{Test Suite}

The implementation is validated through a comprehensive test suite of 631 tests covering:

\textbf{Left-recursion handling:} Direct, indirect, and mutual left recursion; hidden left recursion through sequences and alternatives; left recursion through multiple indirection levels; cache invalidation during LR expansion; interaction between LR and recovery; reference transparency with LR clauses.

\textbf{Error recovery mechanisms:} Character skipping and deletion; recovery at sequence, alternative, and repetition boundaries; bound propagation through nested clauses; error localization and visibility constraints; recovery with partial matches; interaction between recovery and all operator types.

\textbf{Operator correctness:} Sequence composition with complete/incomplete results; first-match (ordered choice) selection with backtracking; repetition (one-or-more, zero-or-more) with recovery; optional clauses; edge cases including empty matches, boundary conditions, and nested structures.

\textbf{Performance and linearity:} Linear time complexity $O(n \cdot |G|)$ verification on inputs up to 100,000 characters; stress tests with deeply nested structures and complex grammars; performance edge cases including pathological backtracking scenarios that would cause exponential behavior in non-packrat parsers.

\textbf{Constraint verification:} Tests explicitly validate all design constraints including phase isolation, monotonic consumption, completeness propagation, LR-recovery separation, parse tree spanning, and visibility.

\textbf{Parse tree spanning invariant:} 20 comprehensive tests verify that all parse results completely span the input from position 0 to end, with total failures returning \texttt{SyntaxError} nodes covering all input, and partial matches wrapped with trailing \texttt{SyntaxError} nodes.

\textbf{Complex interactions:} Combined scenarios testing multiple features simultaneously (e.g., left recursion with error recovery, nested repetitions with boundaries, recovery within LR contexts); unicode handling; real-world parsing patterns.

All tests pass, confirming the algorithm handles corner cases correctly while maintaining linear performance and adhering to the spanning invariant.

\section{Conclusion}

We have presented the squirrel parser, a packrat parser that directly handles left recursion and error recovery while preserving $O(n \cdot |G|)$ time and space complexity. Both extensions are derived from first principles through formal mathematical reasoning.

\textbf{Left recursion.} The algorithm rests on three theorems: fixed-point existence, bottom-up necessity, and monotonic length increase. These necessitate three mechanisms (cycle detection via per-position state tracking, iterative expansion with seed growth, and version-tagged memoization), each proven necessary by contradiction. The implementation requires only a boolean flag per memo entry and a version counter per input position. Unlike prior approaches~\cite{warth2008,tratt2010}, the algorithm handles all recursion types uniformly, requires no grammar preprocessing, and maintains packrat linearity.

\textbf{Error recovery.} We prove that exactly two phases are necessary and sufficient: discovery marks incomplete parses; recovery performs bounded linear search. Phases communicate via a single bit (\texttt{isComplete}), and complete results are phase-independent, enabling cache reuse. The \texttt{isComplete} flag forms a Boolean algebra under conjunction, with PEG operators implementing algebraic operations. Every component is proven necessary by contradiction, establishing algorithmic minimality.

\textbf{Theoretical contributions.} Four deep insights emerged from the derivations: (1) LR expansion implements the Y combinator in imperative form, computing the least fixed point via Kleene iteration; (2) the \texttt{isComplete} flag forms a compositional Boolean algebra, enabling modular reasoning about parser correctness; (3) two-phase parsing is minimal due to existential/universal quantifier separation; (4) a boolean phase flag suffices because only two phases exist.

\textbf{Empirical validation.} The implementation passes over 500 comprehensive tests covering all left-recursion types, error recovery scenarios, operator correctness, performance edge cases, constraint verification, and complex feature interactions. Tests confirm linear performance up to 100,000-character inputs and validate correct handling of pathological cases.

\textbf{Broader implications.} This work demonstrates that neither left recursion nor error recovery fundamentally increases parsing complexity. Both challenges yield to precise algorithmic design grounded in mathematical necessity. The approach generalizes: derive constraints from axioms, construct minimal solutions, prove necessity by contradiction. The resulting algorithms achieve theoretical optimality (every mechanism is necessary, complexity matches the packrat lower bound) while remaining simple enough for straightforward implementation.

The squirrel parser unifies left-recursion handling and error recovery within a single, coherent framework, offering a complete solution for robust PEG parsing without sacrificing the elegance or efficiency that makes packrat parsing attractive.

\bibliographystyle{ACM-Reference-Format}
\bibliography{squirrel_parser}

\end{document}